\documentclass[aps,amsmath,amssymb,superscriptaddress,twocolumn]{revtex4-1}
\usepackage{float}
\usepackage{placeins}
\usepackage{multirow}
\usepackage{booktabs}
\usepackage{comment}
\usepackage{graphicx}
\usepackage{dcolumn}
\usepackage{bm}
\usepackage{layouts}
\usepackage{url}
\usepackage[colorlinks, linkcolor=blue, citecolor=blue, urlcolor=blue,breaklinks=true]{hyperref}

\begin{document}
\title{Surpassing the resistance quantum with a geometric superinductor}
\author{M. Peruzzo}
\email{Authors contributed equally.}
\affiliation{Institute of Science and Technology Austria, 3400 Klosterneuburg, Austria}
\author{A. Trioni}
\email{Authors contributed equally.}
\affiliation{Institute of Science and Technology Austria, 3400 Klosterneuburg, Austria}
\author{F. Hassani}
\affiliation{Institute of Science and Technology Austria, 3400 Klosterneuburg, Austria}
\author{M. Zemlicka}
\affiliation{Institute of Science and Technology Austria, 3400 Klosterneuburg, Austria}
\author{J.~M.~Fink}
\email{jfink@ist.ac.at}
\affiliation{Institute of Science and Technology Austria, 3400 Klosterneuburg, Austria}

\date{\today}

\begin{abstract}
The superconducting circuit community has recently discovered the promising potential of superinductors. These circuit elements have a characteristic impedance exceeding the resistance quantum $R_\text{Q} \approx 6.45\,\text{k}\Omega$ which leads to a suppression of ground state charge fluctuations. Applications include the realization of hardware protected qubits for fault tolerant quantum computing, improved coupling to small dipole moment objects and defining a new quantum metrology standard for the ampere. In this work we refute the widespread notion that superinductors can only be implemented based on kinetic inductance, i.e.\,using disordered superconductors or Josephson junction arrays. We present modeling, fabrication and characterization of 104 planar aluminum coil resonators with a characteristic impedance up to 
30.9\,k$\Omega$ at 5.6\,GHz and a capacitance down to $\leq1$\,fF, with low-loss and a power handling reaching $10^8$ intra-cavity photons. Geometric superinductors are free of uncontrolled tunneling events and offer high reproducibility, linearity and the ability to couple magnetically - properties that significantly broaden the scope of future quantum circuits.
\end{abstract}

\maketitle

\section{Introduction}
In recent years the field of superconducting quantum circuits has been introduced to a new player: the superinductor \cite{kitaev2006,ManucharyanThesis}. It is defined as a circuit element with zero DC resistance and a characteristic impedance $Z_\text{C}$
exceeding the resistance quantum $R_\text{Q} = h/(2e)^2 \approx 6.45\,\text{k}\Omega $. In order for a superinductor to be useful, especially in quantum computing, it needs to have low microwave losses and be in its quantum ground state at millikelvin temperatures. The 
latter requires a self-resonance frequency at least in the GHz range.

Superinductor resonators are particularly interesting for quantum circuits because their impedance affects the local quantum environment. In the ground state, a resonator's 
dimensionless charge and phase fluctuations are impedance dependent and their ratio is given by $\delta n / \delta \phi = R_\text{Q} / Z_\text{C}$.
For this reason superinductors have been identified as necessary to measure the dual element of the Josephson junction, the phase slip junction \cite{Mooij2006}, since the observation of locked phase slip oscillations (dual Shapiro steps) relies on a suppression of charge fluctuations and resistive heating \cite{DiMarco2015,Vora2017,Arndt2018}. Such a measurement gives the prospect of a new quantum metrology standard for current \cite{Likharev1985,Piquemal2000,Flowers2004}. 
In addition, new qubits such as the fluxonium \cite{Manucharyan2009,Grunhaupt2019,Hazard2019} and hardware protected qubits \cite{Doucot2012,Le2019} such as the 0-$\pi$ \cite{kitaev2006,Brooks2013,Groszkowski2018,Paolo2019,gyenis2019}, 
as well as robust quantum error correction schemes \cite{Cohen2017} are dependent on reliable, low-loss superinductors.

High characteristic impedance resonators with fundamental frequency $f_0$ and low stray capacitance $C$ give rise to large zero point voltage fluctuations $V_{\text{ZPF}}=\sqrt{h f_0/(2C)}=2\pi f_0 \sqrt{\hbar Z_\text{C}/2}$. The use of a superinductance can therefore boost conventional coupling limits in circuit QED \cite{Devoret2007,Bosman2017} and facilitate coupling to hybrid piezoelectric components \cite{Arrangoiz_Arriola2018}, as well as to systems having small electric dipole moments, such as electrons in quantum dots 
\cite{Stockklauser2017}
and polar molecules \cite{Andre2006}. Microwave optomechanics experiments such as ground state cooling \cite{Fink2016} and wavelength conversion \cite{Arnold2020} benefit from large $V_{\text{ZPF}}$ but also require parametric coupling, which is increased by a strong coherent drive that relies on a high degree of linearity of the inductance \cite{Peterson2019}.

To date implementations of superinductors are based on kinetic inductance, either by fabricating Josephson junction arrays \cite{Manucharyan2009,Masluk2012,Bell2012,Nguyen2019,Pechenezhskiy2019} or using highly disordered materials, such as nanowires \cite{Peltonen2018,Shearrow2018,Niepce2019,Hazard2019} or granular aluminum \cite{Grunhaupt2018,Grunhaupt2019,Kamenov2020}.
Geometric inductances are widely considered unsuitable as superinductors since reaching $Z_\text{C} > R_\text{Q}$ has been argued to be impossible \cite{Manucharyan2009,ManucharyanThesis,Masluk2012, 
Kamenov2020}. With a simple single-wire resonator the characteristic impedance will indeed be bounded to approximately the impedance of free space $Z_\text{0} = \sqrt{\mu_0/\epsilon_0} = 377\,\Omega$. However, in this work we show that we can exceed this limit by almost two orders of magnitude and satisfy all superinductor requirements. This is achieved by making use of the  mutual inductance contribution of concentric loops in the form of a planar coil together with drastic miniaturization and substrate engineering.

\begin{figure*}[t]
\centering
\includegraphics[width=1\textwidth]{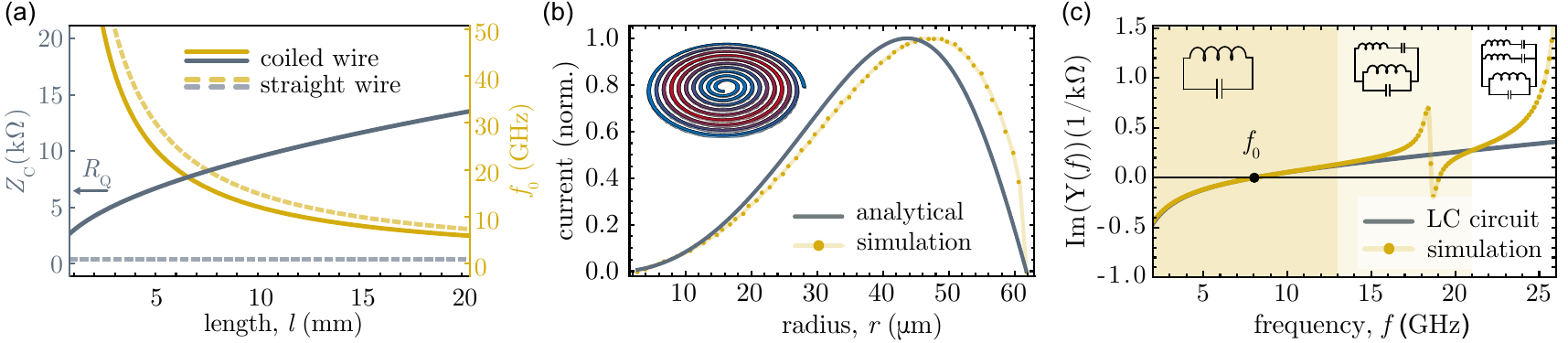}
\caption{
(a) Comparison of characteristic impedance $Z_\text{C}$ (blue) and first resonance frequency (yellow) for a straight wire ($d/w=10$, dashed) and a planar coil ($p=1$\,$\mu$m, $\rho=1$, solid) in vacuum ($\epsilon_\text{eff}=1$). The blue dashed line represents the wire impedance (Eq.\,(\ref{eq:Zwire})), while the blue solid line is the coil characteristic impedance ($Z_\text{C}=2\pi f_g L_g$, with $f_g$ and $L_g$ given by Eqs.\,(\ref{eq:MaleevaFrequency}) and (\ref{eq:Inductance}) respectively). The $Z_\text{C}$ corresponding to coils with lengths $l>4.78$\,mm is above $R_\text{Q} \approx 6.45$\,k$\Omega$, while their resonance frequency (yellow solid line) drops just below 25.4\,GHz. Coil frequency (Eq.\,(\ref{eq:MaleevaFrequency})) and $\lambda/2$ wire resonator frequency $f_0=c_0/2l$, scale similarly with length, with a multiplication factor due to the boosted coil inductance.
(b) Normalized current distribution for a coil with $p=1$\,$\mu$m, $n=60$ and $d_\text{in}=6\,\mu$m. The simulated points (yellow, line is a guide to the eye) follow approximately the analytical model for a $\lambda/2$ resonator (blue). The inset shows the current amplitude distribution where red corresponds to high current values and blue to low ones. As a first approximation, the spiral resonator behaves as a distributed $\lambda/2$ resonator.
(c) Admittance frequency response for the same coil in vacuum. 
The yellow dots are obtained from simulations. The zero crossing (black dot) represents the first resonance frequency. The blue line is the admittance response for a LC circuit with $L_g$ in agreement with Eq.\,(\ref{eq:Inductance}) and the simulated data at low frequencies (dark yellow area). At higher frequencies the admittance displays additional resonance modes, which can be taken into account with the equivalent circuits shown in the insets. The coil can be modeled as an ideal LC circuit around $f_0$, below which it behaves as a pure inductor.
}
\label{fig:coil_modelling}
\end{figure*}

\section{Proof of concept and model}
When constructing a geometric coil inductor it is instructive to compare its characteristic impedance to a distributed wire. The geometric impedance of a wire according to the transmission line model \cite{Pozar2011} is length independent and defined by
\begin{equation}\label{eq:Zwire}
Z_\text{wire} = \frac{1}{\pi}\sqrt{\frac{\mu_0}{\epsilon_0 \epsilon_\text{r}}} \text{arccosh}(d/w), 
\end{equation}
where $w$ is the width of the wire and $d$ is the distance to ground. One could try to rise $Z_{\text{wire}}$ by increasing $d$, since $w$ is ultimately limited by fabrication, by pulling the wire apart from the ground. Nevertheless with a separation $d$ in the order of the wavelength, the circuit starts radiating energy, as if it is shunted by a load resistance in the order of $Z_{\text{0}}$ at resonance \cite{ManucharyanThesis}.
However, by winding the single wire in a planar spiral form the geometric inductance is enhanced as expressed by the current-sheet method formula \cite{Mohan1999}
\begin{equation}\label{eq:Inductance}
L_g = \frac{\mu_0 n^2 d_\text{av} c_1}{2} (\text{ln}(c_2/\rho)+c_3\rho+c_4 \rho^2),
\end{equation}
where $\mu_0$ is the vacuum permeability (assuming none of the material have magnetic properties), $n$ is the number of turns, $d_\text{av} = \frac{d_\text{in} + d_\text{out}}{2}$ is the average between the inner and the outer diameter of the coil and $\rho = \frac{d_\text{out}-d_\text{in}}{d_\text{out}+d_\text{in}}$ is the fill-ratio of the coil, which for our geometries is close to unity. $c_{1,2,3,4}$ are geometry dependent constants, for circular coils considered here they are (1.0, 2.5, 0.0, 0.2). As $d_\text{out}$ is linear in the number of turns, Eq.\,(\ref{eq:Inductance}) shows the geometric inductance increases as $L_g \sim n^3$.

From the analytical model in Ref.\,\cite{Maleeva2015} an expression for the fundamental resonance of a planar circular coil can be calculated as
\begin{equation}\label{eq:MaleevaFrequency}
f_g = \xi\frac{c_0}{\sqrt{\epsilon_\text{eff}}} \frac{2 p}{\pi  (d_\text{in}+2 n p)^2},
\end{equation}
where $\xi$ is a shape-dependant constant which for circular coils is 0.81, $p$ is the pitch which is the distance between adjacent turns (wire width plus spacing), $\epsilon_\text{eff}$ is the effective relative permittivity of the environment, $n$ is the number of turns and $c_0$ is the speed of light in vacuum.
Equation\,(\ref{eq:MaleevaFrequency}) is based on purely geometric considerations and it shows the resonance scaling as $f_g \sim n^{-2} p^{-1} \sim l^{-1}$ \footnote{Approximated expression of the length of a spiral. A spiral can be seen as a series of concentric circles: \unexpanded{$l \simeq \sum_{i=1}^n 2\pi r_i \simeq \sum_{i=1}^n 2\pi i p = 2\pi \frac{n(n+1)}{2} p \simeq n^2 p$}}, where $l$ is the wire length, similar to a distributed element 
resonator.
Modeling the coil as a simple LC oscillator we can express the impedance as $Z_\text{C}=\sqrt{L/C}=2 \pi f_0 L$ and conclude that the characteristic impedance scales with number of turns $Z_\text{C}\sim n$ and that the inductor's parasitic capacitance is linear in the coil radius $C \sim n p$.

The favorable scaling of a coil's impedance and resonance with respect to a straight wire can be seen in Fig.\,\ref{fig:coil_modelling}(a). While the frequency of the wire and the coil scale similarly with respect to length, the impedance of the coil is increasing as $\sqrt{l}/p$. 
The plot shows that the theoretical limit, which constrains the straight wire's characteristic impedance, is lifted for the coil and that the superinductor regime with a fundamental frequency in the GHz range is attainable.
These trends are in quantitative agreement with finite element method (FEM) simulations as
illustrated 
in Figs.\,\ref{fig:coil_modelling}(b) and \ref{fig:coil_modelling}(c).

Figure \ref{fig:coil_modelling}(b) highlights the distributed element behavior of the coil resonator. It displays the current distribution of a coil as a function of radius taken from simulations and that of a $\lambda/2$ resonator given analytically. For the latter the wire is considered to be wound up in a spiral such that the expression for the current can be parameterized by the radius $r$, i.e. $ I(r)=I_0 \sin{(\pi(2r/d_\text{out})}^2)$  \cite{Maleeva2015}. This formula neglects mutual inductance between turns which explains the slight deviation between the two curves.

\begin{figure*}[t]
\centering
\includegraphics[width=1\textwidth]{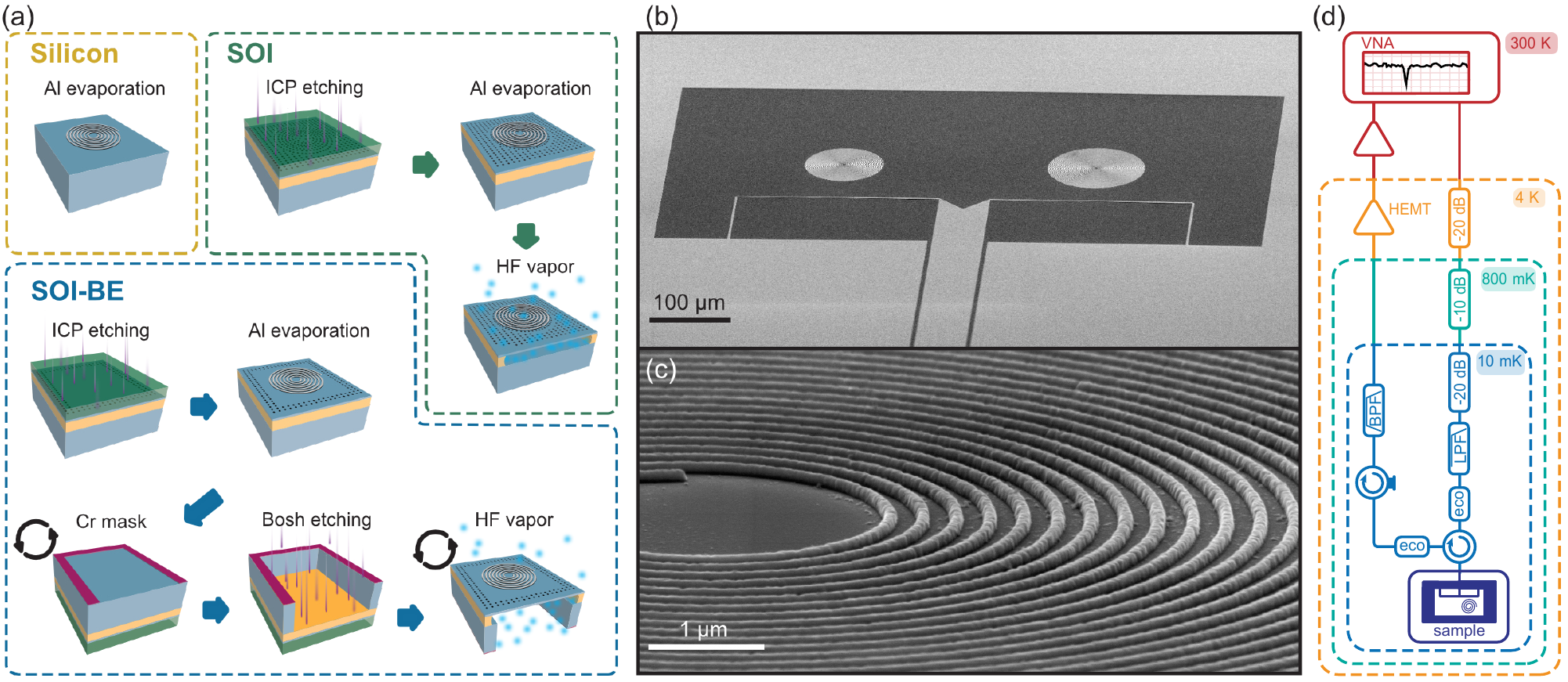}
\caption{
Sample fabrication and measurement setup.
(a) Main nanofabrication steps for each substrate, as described in detail in the Appendix. 
(b) Scanning electron microscopy image of the coil resonators inductively coupled to the shorted coplanar waveguide feedline. 
(c) Enlarged view of a coil resonator with pitch 300\,nm on SOI-BE.
(d) Schematic of the measurement setup. The sample is bonded to a printed circuit board, mounted in a copper sample box and cooled down to 10\,mK in a cryogenic dilution refrigerator. It is shielded by two magnetic shields and a radiation shield. The incoming signal from the vector network analyzer (VNA) is attenuated by approximately 82\,dB (considering attenuators and cable losses) and passes through one low-pass filter (LPF) and a circulator. Two eccosorb filters are attached to the in- and output of the circulator. The outbound signal passes through another circulator for isolation and a band-pass filter (BPF) before being amplified 
by a low temperature (HEMT) and further room temperature amplifiers.}
\label{fig:fab_sem}
\end{figure*}

Figure\,\ref{fig:coil_modelling}(c) compares the simulated frequency dependent admittance $Y$ of a coil shunted with a single lumped port and that of a simple LC circuit, which is $Y(f)=\text{i}\,2\pi f C+1/(\text{i}\,2\pi f L)$.
Following the formalism of black box circuit quantization \cite{Nigg2012}, the first zero crossing represents the fundamental frequency of the coil. The capacitance can be extracted by taking the derivative of the imaginary part of the admittance at that frequency $C= \tfrac{1}{4\pi}\tfrac{\text{d}}{\text{d}f} \text{Im}(Y)|_{f=f_0}$ and we find that the resulting inductance $L_g=\tfrac{1}{C (2\pi f_0)^2}\approx163$\,nH is consistent with Eq.\,(\ref{eq:Inductance}), for the same coil i.e.~$L_g\approx173$\,nH. We get a slightly better agreement, i.e. $L_g\approx180$\,nH, when we extract the resonance frequency $f_0$ and its derivative $\text{d}f_0/\text{d}C$ for the same coil from the simulated $S_{11}$ parameter of a weakly coupled wave port, in direct analogy to the experimental situation shown in Fig.\,\ref{fig:fab_sem}(b). 

The two admittance curves in Fig.\,\ref{fig:coil_modelling}(c) match in the dark shaded area, beyond which adding more LC series circuits to the model increases the accuracy up to the desired frequency (see the inset circuits). This certifies that one can consider the coil as an ideal LC oscillator with characteristic impedance of $Z_\text{C}$ up to and beyond its first resonance. Furthermore, below its fundamental frequency the coil has a negative imaginary admittance, making it an ideal inductor. The claims contained in the rest of this work will pertain only to the frequency region in which the LC approximation is valid.



\section{Fabrication and measurement setup}
It follows from Eqs.\,(\ref{eq:Inductance}) and (\ref{eq:MaleevaFrequency}) and the relation $Z_\text{C}=\sqrt{L/C}=2 \pi f_0 L$ that a coil's impedance can be made larger than $R_\text{Q}$ simply by adding turns, this however decreases the self-resonance which scales as $l^{-1}$. In order to keep the self-resonance in the GHz regime we rely on two tactics: firstly we reduce the pitch to maximize the turn number to length ratio and therefore the impedance per unit length. Secondly we reduce $\epsilon_{\text{eff}}$ of the substrate, which lowers the capacitance without affecting the inductance of the coil.

Regarding the second point we fabricate coils on three different substrates as shown in Fig.\,\ref{fig:fab_sem}(a), i.e.~high resistivity silicon (Si), 220\,nm silicon membrane separated from a silicon handle wafer by 3\,$\mu$m of vacuum (silicon on insulator, SOI), and a fully suspended 220\,nm silicon membrane (SOI backetched, SOI-BE) as extensively described in the Appendix.

For each substrate we fabricate several coils with different turn numbers and pitches and measure their fundamental resonance frequencies in a dilution refrigerator. The coils are coupled to a shorted waveguide, which guarantees inductive coupling as shown in Fig.\,\ref{fig:fab_sem}(b), where we are able to control the coupling strength by adjusting the distance with a resulting extrinsic quality factor $Q_e$ between $5\times10^3$ for a few $\mu$m distance to over $10^5$ for distances around a couple times the coil size. As shown in Fig.\,\ref{fig:fab_sem}(d) we measure the complex $S_{11}$ parameter and fit the I and Q quadrature to a reflective model \cite{Fink2016} to extract external and internal quality factors, $Q_e$ and $Q_i$, and the first resonance $f_0$ as a function of input power.
\begin{figure}[t]
\centering
\includegraphics[width=\columnwidth]{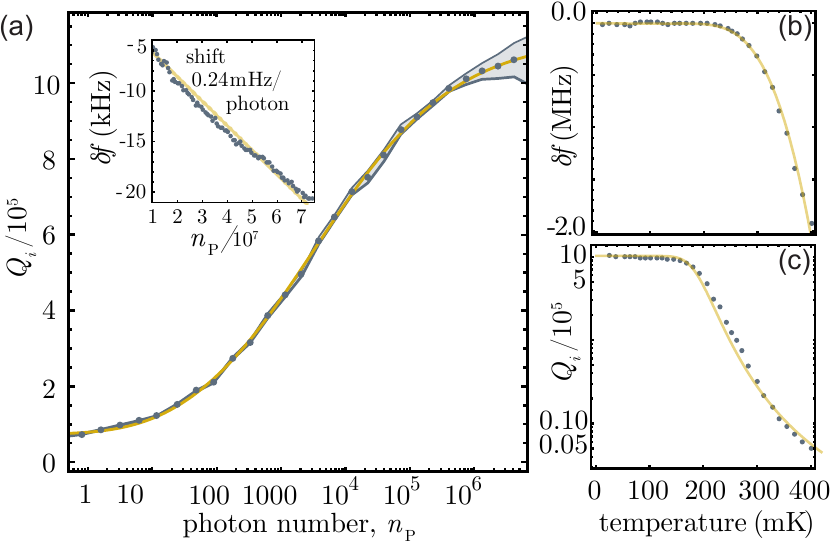}
\caption{
\label{fig:Qs}
(a) Measured internal quality factor $Q_i$ of a representative overcoupled coil ($p=300$\,nm, $n=155$, $Q_e=60 \times 10^3$ on SOI-BE) as a function of the mean intra-resonator photon number $n_\text{P}$. The blue dots are values extracted from a Lorentzian fit and the blue band represents the 90\% confidence interval. The yellow line is the fit to Eq.\,(\ref{eq:TLSfit}), with $F\delta_\text{TLS}$, $\beta$, $n_\text{C}$ and $Q_\text{sat}$ as fit parameters. 
The inset shows the frequency shift $\delta f$ per photon, obtained from a linear fit (yellow) of the measured data (blue dots), resulting in 0.24mHz/photon.
(b)-(c), Temperature sweeps. The blue dots are measured values for the same device as in (a). The yellow lines represent the fits of $\delta f$ and $Q_i$ vs. temperature using Eqs.\,(\ref{eq:BCS_f}) and (\ref{eq:BCS_Q}) respectively. The only fit parameter is $\alpha= 5.9$\% and its small value shows that the inductance is to 94.1\% geometric. 
}
\end{figure}

\section{Loss, kinetic inductance and linearity}
In this section we characterize a reference device on SOI-BE with a pitch of 300\,nm, 155 turns, first resonance at 4.55\,GHz and geometric inductance of 933.9\,nH. It is an average device for the SOI-BE coils that were measured with additional radiation shielding, comparably high $Q_e$ 
and an optimized hydrofluoric acid vapor (VHF) release. Generally speaking, for stronger waveguide coupling we observe lower values of $Q_i$. However, for $Q_e\geq5\times10^4$ the high power $Q_i$ consistently exceeded $10^6$. The full range of values can be found in Table\,\ref{tab:summary}.

Figure\,\ref{fig:Qs}(a) shows the typical dependence of $Q_i$ as a function of the intra-resonator photon number $n_\text{P}$ consistent with the presence of two level systems (TLS) \cite{GaoPhD}
\begin{equation}\label{eq:TLSfit}
Q_\text{TLS}^{-1} \simeq  \frac{F \delta_\text{TLS} \text{tanh}(\hbar \omega/2 k_\text{B}T)}{(1+n_\text{P}/n_\text{C})^\beta}+Q_\text{sat}^{-1},
\end{equation}
where $F$ is the fraction of electric field in the lossy material, $\delta_\text{TLS}$ is the TLS loss tangent, 
$n_\text{C}$ is the critical photon number that saturates the TLS and $Q_\text{sat}=1.1\times 10^6$ represents any additional loss mechanisms. We extract $F \delta_\text{TLS}$ to be 1.3$\times 10^{-5}$, a number that is only about 10 times larger than some of the best reported standard size coplanar aluminum resonators on sapphire \cite{Megrant2012}.
The exponent $\beta$ is commonly taken as 0.5 and deviations indicate an interaction between TLS \cite{Burnett2017}. For our system it was found to be 0.4 which implies some degree of interaction. 

The inset in Fig.\,\ref{fig:Qs}(a) shows the nonlinearity of the resonator, i.e.\,the frequency shift as a function of photon number. From the data we extract a shift of $\delta f=f_0(n_\text{P})-f_0(0) =0.24$\,mHz per photon, a shift so small it requires photon numbers above $10^7$ in order to be meaningfully measured. In general the shifts vary between samples however they are in a similar order of magnitude as the one displayed. This makes these resonators optimal for experiments that require strong parametric driving for enhanced coupling strengths. At extremely high powers the Lorentzian shape of the coil resonance starts to distort indicating a breakdown of superconductivity, this effect occurs at different powers for different coils but consistently arises above $10^6$ photons.

In order to quantify the effect of kinetic inductance we performe temperature sweeps on the same device.
Figures \ref{fig:Qs}(b) and \ref{fig:Qs}(c) show its frequency shift $\delta f (T)=f_0(T)-f_0(0)$ and quality factor degradation $ \delta Q_i^{-1}(T)=Q_i^{-1}(T)-Q_i^{-1}(0)$ as a function of temperature $T$. This behavior can be accurately modeled with BCS theory \cite{Gao2008}
\begin{equation}\label{eq:BCS_f}
     \frac{\delta f (T)}{f_0(0)} = -\frac{\alpha \gamma}{2}\frac{\delta \sigma_2(T,\Delta)}{\sigma_2(T,\Delta)}
\end{equation}
and
\begin{equation}\label{eq:BCS_Q}
     \delta Q_i^{-1} (T) = \alpha \gamma \frac{\delta \sigma_1(T,\Delta)}{\sigma_2(T,\Delta)},
\end{equation}
where $\alpha=L_k/(L_g+L_k)$ is the fraction of the kinetic inductance to total inductance, $\gamma$ is a material dependent parameter, which is -1 for aluminum thin films \cite{GaoPhD}, $\sigma_\text{1}$ and $\sigma_\text{2}$ are the real and imaginary part of the conductance $\sigma = \sigma_1-\text{i}\sigma_2$ as described in \cite{Gao2008}. 
The data is fitted with $\alpha$ as a free parameter while the gap voltage $\Delta$ is taken to be the bulk value for aluminum
\cite{Chubov1969}. 
From the fit we extract $\alpha=5.9\%$, which results in a kinetic inductance of $L_k = 58.0$\,nH. From this value and the following equation
\begin{equation}\label{eq:kinetic_L}
L_k = \mu_0 \lambda_L^2(0) \frac{l}{w h},
\end{equation}
we infer the London penetration depth $\lambda_L(0)$ of 147\,nm \cite{Annunziata2010}, where $\mu_0$ is the vacuum permeability, and $l$, $w$ and $h$ are the length, width and thickness of the wire. This number is henceforth used to estimate the kinetic inductance of all coils to be included in the following analysis. 
The fact that the measured $\lambda_L(0)$ is significantly higher than the value for bulk aluminum (15\,nm \cite{GaoPhD}) is likely due to the thin film nature of the metal \cite{Reale1974}.

\begin{figure}[t]
\centering
\includegraphics[width=1\columnwidth]{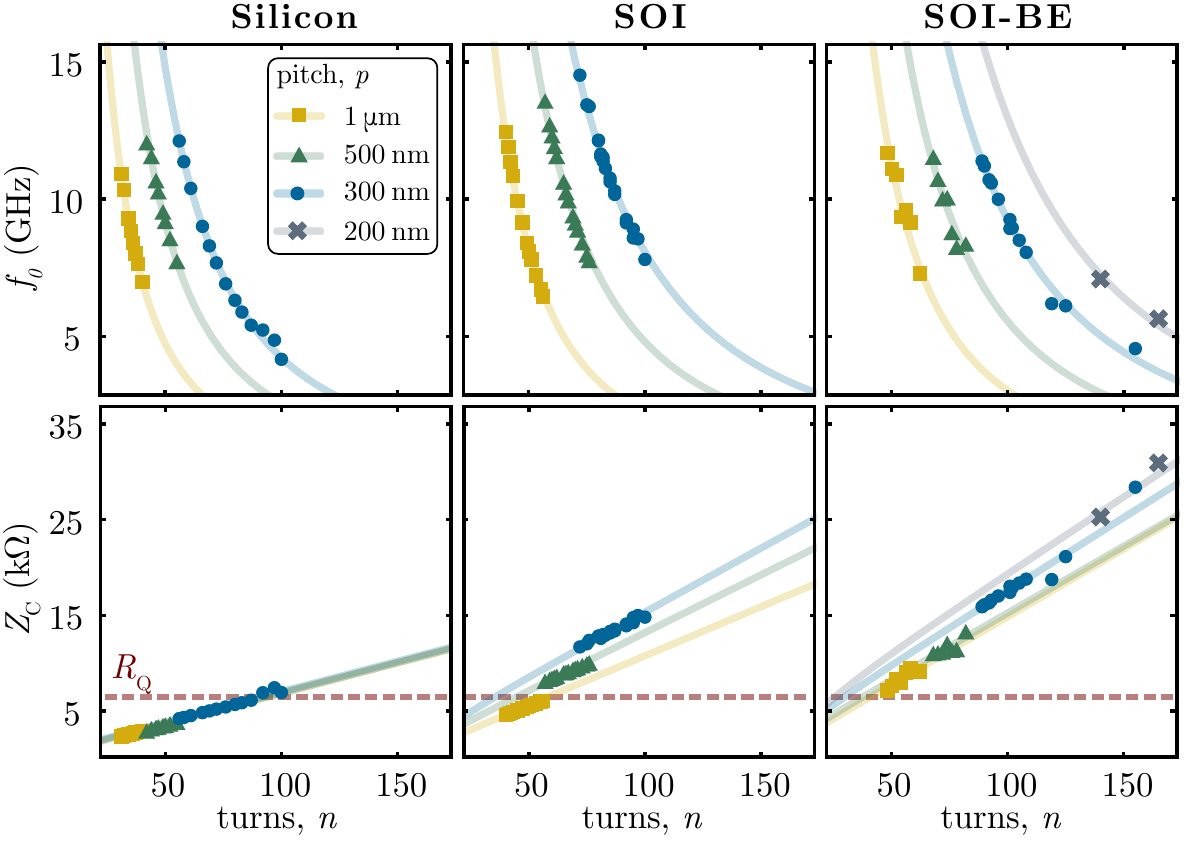}
\caption{
The first row shows all measured resonance frequencies for coils with different pitches $p=1\,\mu$m, 500, 300 and 200\,nm on the three different substrates (Si, SOI and SOI-BE defined in Fig.\,\ref{fig:fab_sem}(a)). The data follow the trend of Eq.\,(\ref{eq:Maleeva_corrected}), with $\epsilon_{\text{eff}}$ as the only fit parameter. 
The second row represents the characteristic impedances calculated with the formula $Z_\text{C}=2\pi f_0 (L_g+L_k)$, where $f_0$ is the measured resonance frequency of each coil and $L_g$ and $L_k$ are from Eqs.\,(\ref{eq:Inductance}) and (\ref{eq:kinetic_L}) respectively. The lines are deduced from the corresponding fit in the first row. While it is possible to surpass $R_\text{Q}$ on silicon, lower permittivity substrates allow for higher frequencies due to a lower coil capacitances.}
\label{fig:setup_data}
\end{figure}

\begin{figure}[t]
\centering
\includegraphics[width=\columnwidth]{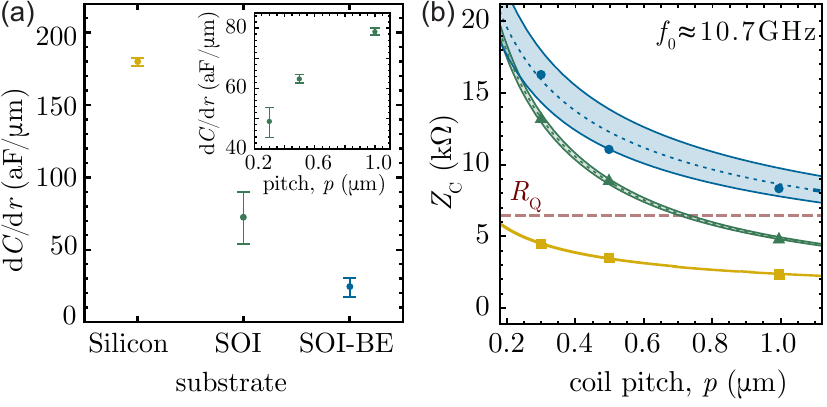}
\caption{
(a) Extracted capacitance per unit radius for three different substrates. The dots represent the average capacitance that would be added by increasing the radius of the coil by one micron. The values are the gradients from linear fits of $C$ vs. radius taken from Fig.\,\ref{fig:setup_data} and averaged over pitch for each substrate. 
The comparable large error bar for SOI results from a pitch dependence shown in the inset. This can be traced back to a pitch dependent participation ratio of the electric field in the silicon handle wafer.
(b) Inferred scaling of the characteristic impedance as a function of pitch at a fixed frequency of $f_0\approx10.7\,$GHz. 
The dashed lines indicate the impedance scaling for each substrate (silicon in yellow, SOI in green, SOI-BE in blue) considering the average $\epsilon_\text{eff}$ extracted from the fits in Fig.\,\ref{fig:setup_data}. The bands account for the permittivity uncertainty from the fit. For SOI the strong pitch dependence of $\epsilon_\text{eff}$ was taken into account with a quadratic interpolation 
for both the bands and the mean values (dashed lines). 
}
\label{fig:eeff_z}
\end{figure}

\section{Characteristic impedance and coil capacitance}
This section summarizes the data extracted from 104 different coils on three substrates. The first row of Fig.\,\ref{fig:setup_data} shows the measured frequency data as a function of the number of turns $n$ for each substrate. The fits (lines) are taken from Eq.\,(\ref{eq:MaleevaFrequency}) with a correction to include the kinetic inductance
\begin{equation}\label{eq:Maleeva_corrected}
f_0 = f_g \sqrt{\frac{L_g}{L_g+L_k}},
\end{equation}
where the only fit parameter is the effective permittivity $\epsilon_\text{eff}$ in $f_g$. The second row in Fig.\,\ref{fig:setup_data} shows the characteristic impedance obtained with $Z_\text{C} = 2\pi f_0 (L_g + L_k)$ where $L_g$ and $L_k$ are calculated with Eqs.\,(\ref{eq:Inductance}) and (\ref{eq:kinetic_L}) and the frequency $f_0$ is given by the data (points) and the fits to the data (lines) shown in the first row. The fits are in excellent agreement with the data points and we find $\epsilon_{\text{eff}}$ of 6.89 $\pm$ 0.09 for silicon, 2.04 $\pm$ 0.93 for SOI and 1.25 $\pm$ 0.19 for SOI-BE. 
In the case of silicon $\epsilon_{\text{eff}}$ can be estimated with 
$\epsilon_{\text{eff}}$ = $(\epsilon_{\text{Si}}+1)/2 = 6.5$, very close to the fit results, and in the case of SOI-BE we find a value close to that of vacuum. The total inductance $L$ is calculated to be in the range from 35 to 992\,nH while the coil capacitance $C$ is in the range 0.88 to 7.71\,fF. A detailed summary can be found in Table\,\ref{tab:summary}. 

The model for $f_g$ in Ref.\,\cite{Maleeva2015} is described as having an error up to 10\%, which could account for some of the variation in the $\epsilon_{\text{eff}}$. Additionally our model does not include parasitic capacitance or mutual inductance due to the coupling wire or the surrounding ground, which in principle can cause small negative and positive shifts in the coil frequency respectively \cite{Hooker2016}. 
Nevertheless, simulations indicate that these coupling related shifts are below 1\%. We are furthermore confident that the measured values correspond to the fundamental mode resonance since wave-port coupled FEM simulations of $f_0$ agree with the data within 20\%.

Figure\,\ref{fig:eeff_z} compares the effects of reducing $\epsilon_\text{eff}$ and pitch on $C$ and $Z_\text{C}$. Reducing $\epsilon_\text{eff}$ has the effect of decreasing  the capacitance, while the inductance remains unchanged as all substrates have $\mu_r=1$. We find that $C$ scales linearly with the coil radius $r \simeq n p$, in agreement with the expected scaling $C\sim n p$ derived earlier. The majority of measured coils had an inner radius of 3\,$\mu$m, however in the limit of large filling factor $\rho$ and $n \gg 1$ the simplification $r = np$ is valid and a coil's self-capacitance can therefore simply be estimated based on its outer radius, highlighting the benefit of miniaturization.

In order to quantify the capacitance suppression we extract the constant gradient $\text{d}C/\text{d}r$ for each pitch $p$ and each substrate resulting in a total of nine data gradients. The extracted gradients are approximately pitch independent and are therefore averaged for each substrate resulting in the three values reported in Fig.\,\ref{fig:eeff_z}(a). The exception is the SOI substrate which exhibits a strong pitch dependence, as shown in the inset of Fig.\,\ref{fig:eeff_z}(a). The reason is that coils with higher pitch have larger size and as a consequence a larger electric field distribution. Because SOI is not a homogeneous material, the higher the pitch the more electric field resides in the silicon handle wafer increasing $\epsilon_\text{eff}$. This is the main reason for the large 90\% confidence interval of $\epsilon_\text{eff}$ reported earlier. 
We observe a larger improvement by changing the substrate from silicon to SOI compared to changing from SOI to SOI-BE 
as can be seen in Figs.\,\ref{fig:eeff_z}(a)-(b). This is because the electric field is more concentrated near the coil and rapidly decays in the vertical direction. Removing the 3\,$\mu$m directly below means placing the strongest part of the field in vacuum which is most effective to lower the effective permittivity. 
Simulations suggest that the capacitance suppression saturates for a vacuum gap of around 20\,$\mu$m depending on the overall coil size.

The data presented in Fig.\,\ref{fig:eeff_z}(b) shows the characteristic impedance for measured coils with similar fundamental frequency $f_0=(10.7\pm0.3)\,$GHz. The superlinear improvement obtained by going to lower pitches occurs because fixing the frequency has the effect of fixing the wire length. For a set length, lower pitch coils have more turns, which gives higher inductance and smaller radii for lower parasitic capacitance. Both have the effect of boosting the characteristic impedance. The dashed lines represent analytical expressions derived from Eqs.\,(\ref{eq:Inductance}), (\ref{eq:kinetic_L}) and (\ref{eq:Maleeva_corrected}) and the bands correspond to the error of $\epsilon_\text{eff}$ as discussed earlier. In the case of SOI the $\epsilon_\text{eff}$ was interpolated between different pitches resulting in a slightly modified shape and a very small error band. Other curves are based on the average value of $\epsilon_\text{eff}$.

\section{Conclusions}
In this study we showed that suspended aluminum coils represent linear low-loss geometric superinductor resonators that can be used as an ideal superinductance below their self-resonance frequency.
For an optimized coupling geometry, fabrication and setup we were able to show $Z_\text{C}\approx 5\times R_\text{Q}$, about 80 times the previously claimed limitation for geometric inductors $Z_0$. It is important to note that, even though the resonators contain a mixture of geometric and kinetic inductance, the geometric component is more than enough to surpass $R_\text{Q}$. In fact by considering the geometric contribution only the highest characteristic impedance becomes 29.62\,k$\Omega$ at $f_0=5.8$\,GHz.

Such a highly miniaturized microwave resonator with large zero point voltage fluctuations reaching $V_\text{ZPF}\approx50\,\mu$V that maintains a linearity of up to $10^8$ photons is an attractive platform for hybrid devices. With losses as low as $Q_i\approx0.8\times10^6$ at single photon powers despite the small gap sizes on the order of 100\,nm, it will also find applications for new quantum circuits. Most importantly, the geometric superinductor is a true single-wavefunction superconducting device with one well defined phase, thus ruling out the possibility of uncontrolled phase and charge tunneling events that could also make it suitable for applications in quantum metrology.


Our study provides simple analytical models and tools to guide future design choices in such applications. Specifically, we found a simple way to predict the coil self-capacitance that only relies on knowing the coil radius and substrate. The design and fabrication method could also help to increase the $Z_\text{C}$ of other superinductors. 
In contrast to circuit elements based on kinetic inductance, we can obtain strong magnetic coupling to feed-lines or other resonators without increasing the parasitic capacitance. 
Together with the flexibility and reproducibility of the geometric inductance this makes them an enabling technology for complex quantum circuits, such as fluxonium and 0-$\pi$ qubits in new error-protected regimes that rely on a high degree of parameter control to avoid symmetry breaking. Potential challenges include increased flux noise due to the large perimeter of the coil \cite{Braumueller2020}, which can be addressed with new materials \cite{place2020} or improved surface fabrication techniques. For example, etching rather than lift-off might enable even smaller coil pitch with better interfaces and lower TLS losses resulting in even higher $Z_\text{C}$.




\section{Acknowledgements}
The authors acknowledge the support from I. Prieto and the IST Nanofabrication Facility. This work was supported by IST Austria and a NOMIS foundation research grant and the Austrian Science Fund (FWF) through BeyondC (F71). MP is the recipient of a P\"ottinger scholarship at IST Austria. JMF acknowledges support from the European Union's Horizon 2020 research and innovation programs under grant agreement No 732894 (FET Proactive HOT), 862644 (FET Open QUARTET), and the European Research Council under grant agreement number 758053 (ERC StG QUNNECT).

\appendix

\section{Device fabrication and parameter summary} \label{appendix}
The devices are fabricated on three different substrates as shown in Figs.\,\ref{fig:fab_sem}(a)-(c) and the results are summarized in Table\,\ref{tab:summary}.

\begin{table*}[t]
\small\centering
\caption{\textbf{Parameter summary of measured geometric superinductor resonators.} $f_0$ are from measurements. $L_g$ and $L_k$ from Eqs.\,(\ref{eq:Inductance}) and (\ref{eq:kinetic_L}) respectively. $C$ and $Z_\text{C}$ are calculated assuming the LC-behavior. $Q_i^\text{low}$ and $Q_i^\text{high}$ are from the fundamental resonance fit at low and high intra-resonator photon numbers. The underlined values refer to devices measured with additional radiation shielding, weaker coupling strengths (high $Q_e$) and optimized VHF release.}
\label{tab:summary}
\resizebox{\textwidth}{!}{
\begin{tabular}{||c|c|c|c|c|c|c|c|c|c|c||} 
			\hline
			Substrate 
			& $p$ [$\mu$m] & diameter [$\mu$m] & $f_0$ [GHz] & $L_g$ [nH] & $L_k$ [nH] & $C$ [fF] & $Z_\text{C}$ [k$\Omega$] & $V_\text{ZPF}$ [$\mu$V] & $Q_i^\text{low} (\times 10^5)$ & $Q_i^\text{high} (\times 10^5)$\\
			\hline\hline
			\midrule 
			\multirow{1}{*}{Silicon} 
			& 1 & 72 - 90 & 10.94 - 7.00 & 32.04 - 62.90 & 2.71 - 4.27 & 6.09 - 7.71 & 2.39 - 2.95 & 24.39 - 17.34  & 0.70 - 1.70 & 1.93 - 15.4 \\
			& 0.5 & 47 - 60 & 12.05 - 7.73 & 35.85 - 74.77 & 4.66 - 7.62 & 4.30 - 5.15 & 2.89 - 3.81 & 30.46 - 22.30 & 0.26 - 1.44 & 3.04 - 17.5 \\
			& 0.3 & 36.6 - 63  & 12.12 - 4.17 & 47.15 - 240.68  & 7.88 - 23.46 & 3.13 - 5.53 & 4.19 - 6.91 & 35.80 - 15.80 & 0.36 - 6.06 & 0.71 - 50.0 \\
			\hline
			\multirow{1}{*}{SOI} 
			& 1 & 86 - 118 & 12.47 - 6.48 & 55.08 - 142.34 & 3.92 - 7.40 & 2.76 - 4.03 & 4.62 - 6.09 & 38.67 - 23.08 & 0.05 - 1.08 & 0.33 - 30.2 \\
			& 0.5 & 63 - 82 & 13.57 - 7.76 & 86.47 - 190.97 & 8.39 - 14.26 & 1.45 - 2.05 & 8.09 - 10.01 & 55.69 - 35.43 & 0.12 - 1.15 & 0.18 - 5.51 \\
			& 0.3 & 49.2 - 66 & 14.52 - 7.80 & 113.97 - 276.27 & 14.13 - 25.59 & 0.94 - 1.38 & 11.68 - 14.80 & 71.59 - 43.31 & 0.02 - 0.37 & 0.22 - 0.82 \\
			\hline
			\multirow{1}{*}{SOI-BE} 
			& 1 & 102 - 130 & 11.71 - 7.31 & 91.93 - 190.32 & 5.53 - 8.99 & 1.90 - 2.38 & 7.17 - 9.15 & 45.22 - 31.88  & 0.16 - 0.51 & 0.20 - 0.92 \\
			& 0.5 & 74 - 88 & 11.53 - 8.36 & 140.29 - 236.08 & 11.60 - 16.44 & 1.25 - 1.43 & 11.00 - 13.27 & 55.17 - 43.95 & 0.08 - 0.71 & 0.27 - 1.73 \\
			& 0.3 & 59.4 - 99 & 11.39 - 4.55 & 201.19 - 933.91 & 20.69 - 57.85 & 0.88 - 1.23 & 15.88 - 28.38 & 65.49 - 35.01 & 0.04 - 0.17  & 0.10 - 0.35  \\
			&&&&&&&&&\underline{0.24 - 1.66}& \underline{1.11 - 10.9}\\
			& 0.2 & 62 - 72 & 7.09 - 5.61 & 515.00 - 807.51 & 50.76 - 68.62 & 0.89 - 0.92 & 25.22 - 30.89 & 51.41 - 44.99 & \underline{1.57 - 7.51} & \underline{4.67 - 21.4} \\
			\hline
\end{tabular}
}
\small 
\end{table*}

In the case of silicon (Si) the fabrication consists of a single layer process. The coil, ground and waveguide are patterned by e-beam lithography (EBL) operated at 100\,keV on a 300\,nm thick layer of CSAR 13 resist. An evaporation of 80 - 100\,nm of aluminum at a rate of 1\,nm/s in ultra-high vacuum (UHV) is followed by a lift-off process in n-metyl-2-pyrrolidine (NMP) at 80$^{\circ}$C resulting in the desired metallic pattern on a high resistivity Si wafer.

The fabrication on silicon-on-insulator (SOI) consists of a multilayer fabrication process. The wafer comprises a 220\,nm thick Si layer on 3\,$\mu$m of silicon dioxide ($\text{SiO}_\text{2}$) which rests on 750\,$\mu$m of Si \cite{Dieterle2016,Keller2017}.  In the first step, EBL is used to pattern arrays of small holes (radius between 65 and 100\,nm depending on pitch) on the thin Si layer around the coil and on the waveguide.
The chip is then mounted in an inductively coupled plasma (ICP) etcher where C$_4$F$_8$ and SF$_6$ etch the holes through the silicon layer reaching the oxide which acts as an etch stop.
Then the coil is patterned and evaporated as in the previous case.
The last step involves vapor hydrofluoric acid (VHF) etching (memsstar ORBIS ALPHA), in which the vapor penetrates through the etched holes and locally removes the oxide underneath, thus terminating the process with a suspended membrane.

The last fabrication routine, aimed to produce SOI-backetched (SOI-BE) samples, is developed using a similar wafer than the former procedure but with the Si handle wafer lapped down to 200\,$\mu$m. The first two steps are identical to the SOI process with the difference that in this case no holes were patterned directly around the coils as they are not needed to suspend the membrane. After the coil metal deposition and consequent lift-off, a layer of 5\,$\mu$m of LOR 5B resist is placed on the structures as protection for the following subprocesses. LOR 5B is chosen specifically because of it does not dissolve in acetone, used in the next lift-off. 
A mask is written in EBL on the back side of the chip into 270\,nm of PMMA 950k EL4 leaving open rectangles positioned directly under the coil.
The rectangles are designed to be large enough for alignment not to be critical. A layer of 50\,nm of chromium (Cr) is deposited and the consequent lift-off carried out by acetone at 40$^{\circ}$C.
The chip is then mounted in an ICP with the devices facing down and the silicon in the rectangles is completely etched away with a customized Bosh process comprised of a gas mixture of C$_4$F$_8$ and SF$_6$. This leaves only the $\text{SiO}_\text{2}$ and 220\,nm silicon layer under the coils.
For the Cr to stick effectively throughout the etch process the bottom of the chip must have a low level of roughness.  The resist is then removed via hot NMP and the oxide layer is locally etched by VHF. Finally the coils are left suspended on a 220\,nm membrane with vacuum below.


\bibliography{Bib}

\end{document}